\documentstyle[12pt]{article}

        \newcommand{\be}{\begin{equation}}
        \newcommand{\ee}{\end{equation}}
        \newcommand{\ba}{\begin{eqnarray}}
        \newcommand{\ea}{\end{eqnarray}}
        \newcommand{\ban}{\begin{eqnarray*}}
        \newcommand{\ean}{\end{eqnarray*}}
        \newcommand{\barr}{\begin{array}}
        \newcommand{\earr}{\end{array}}

        \renewcommand{\H}{{\cal H}}
        \newcommand{\K}{{\cal K}}

        \newcommand{\et}{\hspace{-0.08in}{\bf .}\hspace{0.1in}}
	
        \newcommand{\maps}{\colon}

        \newcommand{\tr}{{\rm tr}}

        \newcommand{\hf}{{1\over 2}}

\def\nn{\nonumber}

\def\qq{\qquad}

\def\g{\gamma}

\def\d{\delta}

\def\e{\epsilon}

\def\S{\Sigma}

\def\pa{\partial}

\def\frac#1#2{{#1\over#2}}

        \newtheorem{ex}{Example}
        \newcommand{\bex}{\begin{ex}\et}
        \newcommand{\eex}{\end{ex}}

        \newcommand{\ed}{\end{document}}

        \def\sqr#1#2{{\vcenter{\vbox{\hrule height.#2pt
                \hbox{\vrule width.#2pt height#1pt \kern#1pt \vrule width.#2pt}
                \hrule height.#2pt}}}}

\font\tenmsb=msbm10 scaled\magstep{1.5}
\newfam\msbfam
\textfont\msbfam=\tenmsb

\newcommand{\A}{{A}}

\newcommand{\lap}{{{}_{{}_{{}_{\sim}}} \!\!\!\!N}}

\newcommand{\R}{{\rm {I\!R}}}

\def\N{{\;\lap (x)\;}}
\def\C{{\cal C}}

\newcounter{figg}

\begin{document}

        \begin{center}
   {\bf Quantum Gravity Hamiltonian for \\
    Manifolds with Boundary \\}
        \vspace{0.5cm}
   {\em John C.\ Baez$\;{}^{1}$, Javier P.\ Muniain$\;{}^{2}$
  and Dardo D.\ P\' {\i}riz$\;{}^{2}$ \\}
   \vspace{0.3cm}
   {\small Departments of ${}^{1}$Mathematics and ${}^{2}$Physics \\
    University of California \\
   Riverside, California 92521 \\}
   \vspace{0.4cm}
   {\small January 16, 1995 \\}
   {\small (Revised version on August 4, 1995) \\}
        \vspace{0.4cm}
        \end{center}

\begin{abstract}
In canonical quantum gravity, when space is a compact manifold with
boundary there is a Hamiltonian given by an integral over the boundary.
Here we compute the action of this `boundary Hamiltonian' on observables
corresponding to open Wilson lines in the new variables formulation of
quantum gravity. In cases where the boundary conditions
fix the metric on the boundary (e.g., in the asymptotically Minkowskian case)
one can obtain a finite result, given by a `shift
operator' generating translations of the Wilson line in the direction of its
tangent vector. A similar shift operator serves as the Hamiltonian constraint
in Morales-T\'ecotl and Rovelli's work on quantum gravity coupled
to Weyl spinors. This suggests the appearance of an induced field theory of
Weyl spinors on the boundary, analogous to that considered in Carlip's work
on the statistical mechanics of the 2+1-dimensional black hole.
\end{abstract}

\subsection*{1 Introduction}

On globally hyperbolic, spatially compact
spacetimes it is a characteristic feature of general relativity that the
Hamiltonian vanishes when Einstein's equations
hold. Suppose that spacetime is of the form $\R\times S$ with $S$ compact
without boundary. Then in the metric representation of general relativity
without matter,
the Hamiltonian density $\H$ is, up to a total divergence,
given by a linear combination of the
components of the Einstein tensor.
The vacuum Einstein equations therefore imply the vanishing of the Hamiltonian
 \be   H_S = \int_S d^3 x\, \lap \H,  \label{1}  \ee
which for this reason is usually called the Hamiltonian constraint.  (Here
$\lap$ is an arbitrary densitized lapse function, and
for simplicity we consider only the case of vanishing shift.)
In canonical quantum gravity one thus expects physical
states to satisfy the Wheeler-DeWitt equation $\hat H_S \Psi = 0$.
This leads to the `problem
of time': the usual recipe for time evolution in quantum mechanics
\[           \Psi \mapsto e^{-it \hat H_S} \Psi, \]
does not capture the dynamics of quantum gravity.
Conceptually, since the state $\Psi$ is
diffeomorphism-invariant, it makes no sense to `evolve $\Psi$ in time'.

Many approaches to this problem have been proposed \cite{Isham}.
A rather obvious strategy is to introduce a nonzero Hamiltonian on
physical states.  Doing so essentially amounts to choosing a
notion of time evolution applicable to physical states.
For example, one can try to gauge-fix Einstein's equation
by using one degree of freedom of the gravitational
field as the time coordinate, so that the canonically conjugate variable
serves as a Hamiltonian generating evolution with respect to  this choice of
time.  Alternatively, one can try to introduce a `clock field': a matter field
whose value serves as as time coordinate, and whose canonically conjugate
field serves as a Hamiltonian density.

Here we consider another way to introduce a nonzero Hamiltonian on physical
states.
Suppose that we take as space a compact manifold $\Sigma$ with boundary
\cite{RT}.
Classically, given initial data on $\Sigma$, Einstein's
equations need not have a unique solution, even locally and up to
diffeomorphism, unless we impose some boundary conditions.  Moreover,
to obtain Hamilton's equations, the Hamiltonian must be functionally
differentiable with respect to the fields on which it depends, at least
with respect to variations preserving the boundary conditions.
Since functional differentiation usually involves an integration by
parts, to obtain a differentiable Hamiltonian one must add a surface term:
\be           H_\S = \int_\S d^3 x\, \lap \H + \int_{\partial \S} d^2S_a\,
\lap \K^a
\label{2} .\ee
The quantity $H_\S$ need not vanish when Einstein's
equations and the boundary conditions hold.  Thus, at least in principle,
we can hope to quantize the theory and obtain a space of states on which
there is a nonzero Hamiltonian.

To what choice of time evolution
does the Hamiltonian $H_\S$ correspond?  The detailed answer, of course,
depends on the choice of boundary conditions and the surface term $\K$.
But in general we may say this: physical states need not be invariant
under diffeomorphisms of spacetime that are not the identity on
$\R \times \partial \Sigma$, and
the Hamiltonian $H_\S$ generates time evolution corresponding to
diffeomorphisms pushing the surface $\{0\} \times \Sigma$ in the
direction $Nv$, where $v$ is the unit timelike vector normal to this
surface.

In what follows
we shall work using Ashtekar's `new variables' \cite{Ash}, namely
a densitized complex triad field $E^a_i$ and a chiral spin connection
$A_b = A_b^j \tau_j$,
where $a,b,c,\dots$ are spacelike indices and $i,j,k,\dots$ are
internal indices.  (Here we omit the tilde sometimes written
over the $E$ field to indicate that it is densitized.)
The Hamiltonian density is then
\[ \H =  - \hf  \e^{ijk} F_{abk} E^a_i E^b_j, \]
and we take as our boundary term
\[ \K^a = \e^{ijk} A_{bk} E^a_i E^b_j .      \]

The resulting Hamiltonian $H_\S$ is compatible with a variety of
boundary conditions.  Smolin \cite{Smolin}, for example, obtains essentially
this
Hamiltonian (but with an additional cosmological constant term) in his study
of quantum gravity with `self-dual'
boundary conditions.  However, these conditions
have not been thoroughly studied yet at the classical level.

Asymptotically Minkowskian boundary conditions
are much better understood, at least classically \cite{Ash}.
Here $\S$ is a ball of coordinate radius $r$, but one is really interested
in the limit as
$r \to \infty$.   The boundary
$\partial \S = S^2$ then represents spacelike infinity, where there is a fixed
Euclidean 3-metric.   In this limit, the triad field
$E_a^i$ is constant on $\partial \S$, and the 3-metric at spacelike infinity
is given by $ q q_{ab} = \delta_{ij} E_a^i E_b^j$.
When we set $\lap = 1$,
the Hamiltonian $H_\S$ then generates time evolution with respect to the
standard Minkowski time coordinate at spacelike infinity.
One might hope, therefore, that the corresponding quantum
Hamiltonian $\hat H_\S$ will generate
nontrivial time evolution for asymptotically Minkowskian states of
{\it quantum}
gravity.  Of course the very notion of an asymptotically flat state of
quantum gravity is
problematic.  However, the extent of the problems can only be understood by
investigation.
%It is worth comparing the idea that interacting quantum fields
%approach free fields as $t \to
%\pm \infty$; while this  is fundamentally erroneous, as witnessed by
%Haag's no-go theorem, it is nonetheless very useful in practical computations.

%Skip this figure!!!
%\figura{ball.eps}{2}{Compact space $S$ split as
%$\S \cup_{\partial \S} \S '$}

In most of  what follows we will not need a specific choice of boundary
conditions;
all we will need is boundary conditions for which $E^a_i$ and $A_b^j$
have the usual Poisson brackets $\{E^a_i(x), A_b^j(y)\}
= -i\delta^a_b \delta_i^j \delta^3(x,y)$.
This is not true for self-dual boundary conditions, where there are extra
boundary terms for the Poisson brackets, but it is true in the
asymptotically
flat case. Given these Poisson brackets,
one then expects the corresponding quantum operators to be
\ban  && \hat \H = - \hf  \e^{ijk} F_{abk} {\d\over{\d A_a^i}}
{\d\over{\d A_b^j}},   \\
&& \hat \K^a = \e^{ijk} A_{bk} {\d\over{\d A_a^i}}{\d\over{\d A_b^j}}  \ean
in units where $\hbar = 1$.

Here we should note that two operator
orderings for the Hamiltonian constraint have been widely studied,
the `$FEE$' and `$EEF$' orderings, and the former turns out
to be the appropriate one
when we think of the constraint as acting on
wavefunctions on the space of connections (rather than
dually on measures on the space of connections).  Briefly, the $FEE$
ordering is the one for which the constraint may be described as
a `shift operator' \cite{RovSmo},
and also the one for which the relation between
quantum gravity with cosmological constant and Chern-Simons
theory becomes apparent \cite{Baez}.  Thus we adopt this ordering and
the compatible ordering for the boundary term.

Crucial to Rovelli and Smolin's \cite{RovSmo}
original paper on the loop representation
of quantum gravity was their calculation of the action of the
Hamiltonian constraint on Wilson loops.
Surprisingly, they obtained a {\it 3-dimensional} geometrical
interpretation of the Hamiltonian constraint, which allowed them
to find --- at least heuristically --- a large set of solutions of the
Wheeler-DeWitt equation.

Recall that in this work, Wilson loops play a dual role.
In the Heisenberg picture we may think of them as
multiplication operators of the form $\tr(U[1,0])$, where
\[  U[t',t] = {\cal P}\;\exp \int_t^{t'} ds\,{\dot\g}^c (s)\,
A^k_c(\g(s))\,\tau_k   \]
is the holonomy from $t$ to $t'$ and $\g \maps [0,1] \to S$ is a loop.
These form a maximal commuting set
of gauge-invariant kinematical observables, on which the
diffeomorphism constraint acts in a simple way.
Alternatively, in the Schr\"odinger picture we may think of Wilson loops
as kinematical states, that is, wavefunctions
of the form $ \Psi(A) = \tr(U[1,0])$ on the space of connections.
This permits the construction of a loop representation in which
states solving the
diffeomorphism constraint are described by (diffeomorphism equivalence
classes of) collections of loops.

In the Heisenberg picture
the action of the Hamiltonian constraint on Wilson loops
is given, as usual, by a commutator
\[  \Big[ \hat H_S, \tr(U[1,0]) \Big], \]
while in the Schr\"odinger picture it is given by
\[    \hat H_S\,\Psi .\]
In either case, a regularization procedure is needed to
compute the action.  Initially, Rovelli and Smolin used a simple
point-splitting
regularization, and obtained an integral involving $F_{ab}^i \dot\gamma^a
\dot \gamma^b$.
Since $F_{ab}^i$ is antisymmetric in the indices $a,b$ this vanishes
when $\g$ is smooth and without self-intersections.   They
concluded that the Wilson loop states associated to such loops
were solutions of the Wheeler-DeWitt equation, and thus physical states.

Later work made it clearer that the regularization
issues are quite delicate \cite{Blencowe,Borissov,BP,G,JacSmo}.
In fact, they remain controversial, and mathematically rigorous work on the
loop representation is just nearing the point of being able to definitively
deal with them \cite{Baez,Ash2}.   We will not address
these issues in the present work.  Instead, we will work at a level of rigor
similar to that of Rovelli and Smolin's original work, and concentrate on the
new features that arise when space is a manifold with boundary.

\subsection*{2 Hamiltonian Action on Wilson Lines}

%(See Fig.\ 2.)
%%%%%%%%%%%%%%%%%%%%%%%%%%%%%
%Here we want a figure of a ball labelled \S with a Wilson
%  line in it,
%perhaps labelled \g, and perhaps with endpoints
%  \g_0 and \g_1 labelled
%%%%%%%%%%%%%%%%%%%%%%%%%%%%%%%
%\figura{S2.eps}{3}{Wilson line}

When space has no boundary, physical states of quantum gravity in terms of
the new variables are invariant under ${\rm SL}(2,\C)$ gauge transformations.
When space is a compact manifold $\Sigma$
with boundary, the boundary
conditions may break gauge-invariance at the boundary.   For example,
in the asymptotically Minkowskian case the condition that $E^a_i$ be constant
at the boundary is not gauge-invariant.   In such cases, Wilson loops will not
suffice as a complete set of kinematical states in the Schr\"odinger picture,
since loop (or multi-loop) states are gauge-invariant.  Similarly, in the
Heisenberg
picture the Wilson loops will not form a complete (i.e., maximal commuting)
set of kinematical observables.

For this reason it is important to
consider the action of the Hamiltonian not only on
Wilson loops but also on `Wilson lines' starting and ending on the boundary.
Let $\g$ be a smooth path for which the initial and final points $\g_0 =
\g(0)$ and
$\g_1 = \g(1)$ lie on $\partial \S$.  Then in the Heisenberg picture,
the associated Wilson line is the matrix-valued observable
$U[1,0]$.  Now recall that $A_b$ is a chiral spin
connection, so the holonomy $U[1,0]$ describes the parallel
transport of Weyl spinors along the curve $\g$.
If we fix Weyl spinors  $\psi$ and $\psi'$
at the endpoints of $\g$, then
in the Schr\"odinger picture we may also define a Wilson line state by
\be    \Psi(A)  = \psi_A '\; U[1,0]^A_B \; \psi^B.  \label{linestate} \ee
These Wilson lines are invariant under gauge
transformations that are the identity on the boundary.
Moreover, by using Wilson lines in addition
to Wilson loops, one can obtain
a complete set of kinematical observables (or states) that are invariant
under gauge
transformations that are the identity on the boundary \cite{Baez,Ash2}.

To the same degree of
rigor as in the case without boundary,  Rovelli and Smolin's
argument shows that the action of the Hamiltonian on smooth Wilson
loops without self-intersections is identically zero.
The simplest case exhibiting the effect of the boundary term is that of a
smooth Wilson line $\g$ intersecting the boundary transversally at its
endpoints $\g_0$, $\g_1$.   Since
\[         \left[ \int_\S d^3 x \, \lap \hat \H, U[1,0] \right] = 0\]
by Rovelli and Smolin's original computation, it follows that
\[         \left[\hat H_\S, U[1,0] \right] = \left[\hat
H_{\pa \S},  U[1,0]\right].    \]
Using the fact that
\[    \left[{\d\over{\d A_a^i(x)}}\;{\d\over{\d A_b^j(x)}} , U[1,0]\right] =
{{\d^2 U[1,0]}\over{\d A_a^i(x)\;\d A_b^j(x)}} +
{{\d U[1,0]} \over{\d A_a^i(x)}}\,{\d \over{\d A_b^j(x)}} +
{{\d U[1,0]}\over{\d A_b^j(x)}}\,{\d \over{\d A_a^i(x)}},   \]
it follows that
\[         \left[\hat H_\S, U[1,0] \right] = \C_1 + \C_2 + \C_3  \]
where, letting $A^{ij}_b =\e^{ijk} A_{bk}$, we have
\ban      \C_1 &=& \int_{\pa\S} d^2 S_a (x)\lap(x)  A^{ij}_b (x)\; {\d\over{\d
A_a^i(x)}}\;{\d\over{\d A_b^j(x)}}\;U[1,0],     \\
\C_2 &=&  \int_{\pa\S} d^2 S_a (x)\lap(x)
A^{ij}_b (x)\; {\d U[1,0] \over{\d A_a^i(x)}}\;{\d\over{\d A_b^j(x)}},   \\
\C_3 &=& \int_{\pa\S} d^2 S_a (x)\lap(x)  A^{ij}_b (x)\;
{\d U[1,0] \over{\d A_b^j(x)}}\;{\d\over{\d A_a^i(x)}}.
\ean
If instead we work in the Schr\"odinger picture and define a Wilson
line state by eq.\ (\ref{linestate}), we have
\be    \hat H_\S \; \Psi = \psi_A '\; \left[\hat H_\S, U[1,0] \right] \;
\psi^B  = \psi_A '\; {\C_1}^A_B \; \psi^B.   \label{Schr} \ee

In what follows we compute $\C_1$, $\C_2$ and $\C_3$,
with the results appearing in eqs.\ (\ref{5}), (\ref{6}) and  (\ref{7}),
respectively.
We begin by evaluating $\C_1$, which can be
written as an integral over $\S$ of a total divergence:
\[      \C_1 = \int_{\S} d^3 x \;\pa^x_a\;\left(\N A^{ij}_b (x)\; {\d\over{\d
A_a^i(x)}}\;{\d\over{\d A_b^j(x)}}\;U[1,0]\right).    \]
Introducing a point-splitting by letting $z_\e(x,y)$ be a function that
tends to $\d^3(x,y)$ as $\e \downarrow 0$, we have
\[    \C_1 =
\int_{\S} d^3 x \;\pa^x_a\;\left(\int_{\S} d^3 y\, z_\e(x,y)  \N A^{ij}_b (y)\;
{\d\over{\d A_a^i(x)}}\;{\d\over{\d A_b^j (y)}}\;U[1,0] \right).       \]
 Note that the reason we rewrite $\C_1$ as an integral over $\S$ is
 precisely to carry out this point-splitting.

Since
\ban
&&{\d\over{\d A_a^i(x)}}\;{\d\over{\d A_b^j(y)}}\;U[1,0] = \int_0^1 ds \int_0^1
dt\;{\dot\g}^a (s)\; {\dot\g}^b (t)\; \d(x,\g(s))\;\d(y,\g(t))\\
&&\left(\theta(t-s)\; U[1,t]\,\tau_i\,U[t,s]\,\tau_j\, U[s,0] + \theta(s-t)\;
U[1,s]\,\tau_j\,U[s,t]\,\tau_i\, U[t,0]\right),
\ean
we obtain
\ban
\C_1 &=&
\int_0^1 ds \int_0^1 dt\;\int d^3 x\; \int d^3 y\; \pa^x_a\;\Big( z_\e(x,y) \N
A^{ij}_b (y)\lbrace{\dot\g}^a (s)\;  {\dot\g}^b (t)\;  \\
&& \d(x,\g(s)) \d(y,\g(t)) -
{\dot\g}^a (t)\; {\dot\g}^b (s)\; \d(x,\g(t))  \d(y,\g(s))\rbrace\Big)\;
T_{ij}(t,s),
\ean
where
\[ T_{ij} (t,s) = \theta (t-s) U[1,t]\,\tau_i\,U[t,s]\,\tau_j\, U[s,0].   \]
Turning this back into a
surface integral and doing the integral over $y$,  this gives

\ban
\C_1 &=& \int_0^1 ds \int_0^1 dt \int_{\pa\S}
d^2 S_a(x) \N \Big( z_\e(x,\g(t)) A^{ij}_b(\g(t))
{\dot\g}^a (s) {\dot\g}^b (t) \d(x,\g(s)) -    \\
&& \qq \qq z_\e(x,\g(s)) A^{ij}_b (\g(s)) {\dot\g}^a (t) {\dot\g}^b (s)
\d(x,\g(t))\Big)\;T_{ij}(t,s).
\ean
Using the fact that for $x \in\pa\S,$
\be
\int_0^1 ds {\dot\g}^a (s)\int d^2 S_a(x)\;\d(x,\g(s))\;f(x) = f(\g_0) +
f(\g_1),
\label{jacobian}
\ee
where $d^2 S_a (x) = d^2 x \, n_a (x)$, we obtain
\ban
\C_1 &=&
\int_0^1 dt\;\lap (\g_0) z_\e(\g_0,\g(t))\; A^{ij}_b (\g(t))\; {\dot\g}^b (t)
T_{ij}(t,0) -   \\
&& \int_0^1 dt\;\lap (\g_1) z_\e(\g_1,\g(t))\;
A^{ij}_b (\g(t))\;{\dot\g}^b (t)\; T_{ij} (1,t).
\ean
Taking the limit as $\e\to 0$, and using the fact that $\g_0 \not= \g_1$,  we
have
\[    \lim_{\e\to 0} z_\e(\g_0,\g(t))= \d^{(2)} (\g_0,\g_0)\;{{\d (t)}
\over{n_c(\g_0) {\dot\g}^c(0)}},         \]
and a similar result for $z_\e(\g_1,\g(t))$. Thus we obtain
\ban
\C_1 &=& \lap (\g_0) \d^{(2)} (\g_0,\g_0)\;{{A^{ij}_b (\g_0)\, {\dot\g}^b (0)}
\over{n_c
(\g_0) {\dot\g}^c(0)}}\;U[1,0] \tau_i \tau_j -  \\ && \lap (\g_1) \d^{(2)}
(\g_1,\g_1)\;{{A^{ij}_b (\g_1)\,
{\dot\g}^b (1)}\over{n_c (\g_1) {\dot\g}^c(1)}}\;\tau_i \tau_j U[1,0]. \ean
Using the identity $\e^{ijk} \tau_i \tau_j = 2i\;\tau^k$, we obtain the final
result:
\ba
\C_1 &=& 2i\; \lap (\g_0) \d^{(2)} (\g_0,\g_0)\; U[1,0]
{{\A_b (\g_0)\,{\dot\g}^b (0)}\over{n_c (\g_0) {\dot\g}^c(0)}} - \nn \\
&& 2i \;\lap (\g_1)\d^{(2)} (\g_1,\g_1)\;
{{\A_b (\g_1)\,{\dot\g}^b (1)} \over{n_c (\g_1) {\dot\g}^c(1)}}\;
U[1,0],  \label{5}
\ea
where we have defined $\A_b = A_b^k \;\tau_k$.
Note that this result is purely formal, as $\d^{(2)} (\g_0,\g_0)$ is infinite.
It should be regarded as a precise
description of the behavior of $\C_1$ in the limit as
the point-splitting function $z_\e (x,y)$ converges to $\d (x,y)$.
(At the end of this section we
discuss how to `renormalize' the Hamiltonian to obtain a finite
result.)

In a similar manner we can evaluate the second term of
$ \Big[ H_{\pa\S}, U[1,0] \Big] $, obtaining
\ban
\C_2 &=& \int_{\S} d^3 x \pa^x_a \Big( \N  A^{ij}_b (x)
{{\d U[1,0]}\over{\d A_a^i(x)}} {\d\over{\d A_b^j(x)}}\Big)        \\
&=& \int_0^1 dt
\int_{\S} d^3 x \pa^x_a  \Big( {\dot\g}^a (t)   \N A^{ij}_b (x) \d(x,\g(t))
U[1,t]
\tau_i\;U[t,0] {\d\over{\d A_b^j(x)}}\Big)        \\
&=& \int_0^1 dt {\dot\g}^a (t) \int_{\pa\S} d^2 S_a (x)
 \d(x,\g(t)) \N A^{ij}_b (x) U[1,t]\tau_i U[t,0] {\d\over{\d A_b^j(x)}}. \ean
By (\ref{jacobian}), we obtain:
\ba
\C_2 &=& \lap(\g_0) \e^{ijk} A_{bk} (\g_0)  U[1,0]  \tau_i
{\d\over{\d A_b^j(\g_0)}} +
\nn   \\
&& \lap(\g_1)  \e^{ijk} A_{bk} (\g_1) \tau_i U[1,0]
{\d\over{\d A_b^j(\g_1)}}.    \label{6}
\ea
With the help of some Pauli matrix identities, and setting
\[    {\d\over{\d \A_a}}\; = \; \tau_i\; {\d\over{\d A_a^i}}, \]
the final result can be written as
\[  \C_2 \;\psi =
i\;\lap (\g_0)\;U[1,0]\; \Big[ \A_b (\g_0), {{\d\psi}\over {\d \A_b (\g_0)}}
\Big] +
i\;\lap(\g_1)\; \Big[ \A_b (\g_1), {{\d\psi} \over{\d \A_b (\g_1)}} \Big]\;
U[1,0]. \]

Proceeding in the same way as before we can evaluate the third term in the
commutator as
\ba
\C_3 \;\psi &=& i\;\lap (\g_0)\;U[1,0]\;
\Big[ n_a(\g_0) {{\d\psi}\over{\d \A_a (\g_0)}},
{{\A_b (\g_0) {\dot\g}^b(0)}\over{n_c (\g_0) {\dot\g}^c(0)}} \Big] +  \nn
\\ && i\;\lap (\g_1)\;
\Big[ n_a(\g_1) {{\d\psi}\over{\d \A_a (\g_1)}},
{{\A_b (\g_1) {\dot\g}^b(1)}\over{n_c (\g_1) {\dot\g}^c(1)}} \Big]\;U[1,0].
\label{7} \ea

It is unfortunate, but not unexpected, that the term
$\C_1$ which determines the action of the Hamiltonian
on a Wilson line state is singular.  In fact,
the action of the Hamiltonian constraint on Wilson loop states is singular in
a very similar way.  Rovelli and Smolin dealt with this problem
by point-splitting the Hamiltonian constraint
and then `renormalizing' it, that is, multiplying it by $\e$ before taking the
limit as $\e \to 0$.  We can do the same sort of thing when our boundary
conditions are such that the 3-metric is fixed on $\pa \S$ --- for example,
in the asymptotically flat case.
Namely, we pick any metric $q_{ij}$ on $\S$ extending
the fixed metric on $\pa \S$, and using this metric choose the regulator to be
\[         z_\e(x,y) = \sqrt{q/{(\pi\e)}^3} e^{-d(x,y)^2/ \e}, \]
where $d(x,y)^2 \sim q_{ij}(x-y)^i(x-y)^j$ is the square of the distance
{}from $x$ to $y$, and $q$ stands for the determinant of the metric.
We then renormalize the Hamiltonian $\H_\S(\e)$ by writing the boundary term
as an integral of a total divergence, performing a point-splitting
of both terms, and introducing a factor of $\e$ before taking the $\e \to 0$
limit:
\ban   \hat \H_\S^{ren} &=&
 \lim_{\e \to 0}  \e
\big\{-\hf \int_\S d^3 x \;z_\e(x,y)
\N \e^{ijk} F_{abk}(x) {\d\over{\d A_a^i(x)}} {\d\over{\d A_b^j(y)}} + \\
&&
\int_{\S} d^3 x \;\pa^x_a \left(\int_{\S} d^3 y\, z_\e(x,y)  \N A^{ij}_b (y)\,
{\d\over{\d A_a^i(x)}}\,{\d\over{\d A_b^j (y)}}\right) \big\}  .\ean
If we apply this to the Wilson line state given in eq.\ (\ref{linestate}),
Rovelli and Smolin's argument shows that the first term is zero.
Following our earlier computation of $\C_1$ but using the formula
\[    \lim_{\e \to 0} \e z_\e(\g_0,\g(t))
=  {{\d (t)} \over{n_c (\g_0){\dot\g}^c(0)}},         \]
we find that the second term gives
\ba    \hat H_\S^{ren} \; \Psi &=&
2i \psi \big( \lap (\g_0) \; U[1,0]
{{\A_b (\g_0)\,{\dot\g}^b (0)}\over{n_c (\g_0) {\dot\g}^c(0)}} - \nn \\
&& \lap (\g_1) \;
{{\A_b (\g_1)\,{\dot\g}^b (1)} \over{n_c (\g_1) {\dot\g}^c(1)}}\;
U[1,0] \big) \psi,  \label{final}
\ea

Now one might worry that since our regulator depended on a choice of
metric $q_{ij}$ on $\S$, this renormalization prescription spoils
the diffeomorphism-invariance of the problem.  However,
we are assuming that the boundary conditions are such that the metric
$q_{ij}$ is fixed on $\pa \S$, so the symmetry group has been reduced to the
group of diffeomorphisms that preserve this metric on $\pa \S$.  Note that
the final result in eq.\ (\ref{final}) only depends on the metric via
the unit normal vector $n_c$.  It thus depends only on the value of $q_{ij}$
on $\pa \S$.  Thus $\H_\S^{ren}$ is preserved, as it
should be, by all diffeomorphisms of $\S$ preserving the metric on $\pa \S$.

\subsection*{3 Conclusions}

In addition to its conceptual subtleties, canonical quantum gravity presents
many technical problems, so the regularization procedure leading to
our final result should be carefully checked.  It is encouraging,
however, that eq.\ (\ref{final}) has a strikingly simple geometrical
interpretation.
The Hamiltonian $\hat H_\S^{ren}$ acts on the Wilson line state $\Psi$
to give a difference of two terms.  In the first, $U[1,0]$ is multiplied on
the right by a
term proportional to the component of the connection $A$ in the direction
${\dot\g} (0)$.  In the second, $U[1,0]$ is multiplied on the left by a term
proportional
to the component of $A$ in the direction ${\dot\g} (1)$.  In short, the
Hamiltonian acts as a kind of `shift operator' corresponding to an
infinitesimal
displacement of the Wilson line in the direction of its tangent vector.

This `shift operator' interpretation is closely related to
two earlier results in the loop representation of quantum gravity.   First,
Rovelli
and Smolin \cite{RovSmo} have already shown that in the case of a space
without boundary, the Hamiltonian constraint acts on Wilson loops as a
shift operator.  This 3-dimensional geometrical interpretation of the
Hamiltonian constraint is what enabled them (and others) to find solutions of
the Wheeler-DeWitt equation.

Second, a formula very similar to ours also arises in
the work by Morales-T\'ecotl and Rovelli on quantum gravity coupled to
massless chiral fermions, that is, a Weyl
spinor field \cite{TecRov}.  They consider a space without boundary, so there
is no Hamiltonian, only a Hamiltonian constraint.  Moreover, while for us
the spinors $\psi^A$
appearing at the ends of Wilson lines are only a device to extract numbers out
of the matrix-valued observable $U[1,0]$, for them the spinors are dynamical
fields appearing in the Lagrangian.  Nonetheless, when they switch to the
Schr\"odinger representation, construct states as in eq.\ (\ref{linestate})
and compute the action of the Hamiltonian constraint on these states, the
answer is given by a shift operator as in our work.

Why should the Hamiltonian for quantum gravity
on a space with boundary have the same action on Wilson lines
as the Hamiltonian constraint for quantum gravity
coupled to Weyl spinors?  A clue is provided by Carlip's computation
\cite{carlip}
of black hole entropy in 2+1-dimensional quantum gravity.  He treats the event
horizon as a boundary and notes, as we do above, that the boundary conditions
break diffeomorphism invariance at the boundary.  Due to this reduction of
symmetry, modes of the gravitational field that would otherwise be treated as
`gauge' manifest themselves as physical degrees of freedom.    It is these
degrees of freedom that account for the black hole entropy in his computation.

Technically speaking, Carlip proceeds via Witten's description of
2+1 quantum gravity (with cosmological term) as a Chern-Simons
theory, and uses the fact that Chern-Simons theory on a manifold with
boundary induces a field theory on the boundary, namely a WZW model.
As noted by Balachandran {\it et al} \cite{BCM}, the mathematics involved
here is the same as that which describes the fractional quantum Hall effect.
In the fractional quantum Hall effect the induced WZW model is known
to describe a chiral fermion field on the boundary.

Balachandran {\it et al} have
suggested that in 3+1-dimensional quantum
gravity as well, boundary conditions breaking diffeomorphism invariance
should give rise to an induced field theory on the boundary, whose `edge
states'
account for the black hole entropy. Our work suggests that when the
boundary conditions fix the metric at the boundary, this induced field theory
describes a Weyl spinor field coupled to the connection $A_b$ by means of
the Hamiltonian given in eq.\ (\ref{final}).  In this interpretation the
endpointsof Wilson lines act as chiral fermions living on $\pa \S$.
Trying to work out the black hole entropy by this method is
especially tempting, because the area operator in the loop representation
essentially counts the number
points where Wilson lines intersect a surface \cite{RS}.  Smolin has already
begun work towards proving area-entropy relations for black holes
by exploiting this fact \cite{Smolin}.   It is also interesting to
compare the work of Hawking and Horowitz \cite{HH}.

\subsection*{Acknowledgments}

We would like to thank Abhay Ashtekar,
Steve Carlip, Carlo Rovelli, and Lee Smolin for helpful discussions.
D.\ P.\ acknowledges CONICYT
(Uruguay) for partial support.

\vfill

\ed

--------------------------------------------------------------
Javier P. Muniain                     |   Phone (909) 787-7356
Department of Physics
University of California, Riverside
--------------------------------------------------------------
\begin{thebibliography}{10}

\bibitem{Isham} C.\ Isham, Canonical quantum gravity and the problem of time,
in {\sl Integral Systems, Quantum Groups, and Quantum Field Theories,}
eds.\ L.\ A.\ Ibort and M.\
A.\ Rodriguez, Kluwer, Dordrecht, 1993, pp.\ 157-207.

\bibitem{RT} T.\ Regge and C.\ Teitelboim, Role of surface integrals in the
Hamiltonian formulation of general relativity, {\sl Ann.\ Phys.\ }{\bf 88}
(1974), 286-318.

\bibitem{Ash} A.\ Ashtekar, New hamiltonian formulation of general relativity,
{\sl Phys.\ Rev.\ }{\bf D36} (1987), 1587-1602.

A.\ Ashtekar and invited contributors, {\sl New Perspectives in Canonical
Gravity,} Bibliopolis, Napoli, 1988.

\bibitem{Smolin} L.\ Smolin, Linking topological quantum field theory and
nonperturbative quantum gravity, Report CGPG-95/4-5, preprint
available as gr-qc/9505028.

\bibitem{RovSmo} C.\ Rovelli and L.\ Smolin, Loop representation for
quantum general relativity, {\sl Nucl.\ Phys.\ }{\bf B331} (1990), 80-152.

\bibitem{Baez} J.\ Baez, Knots and quantum gravity: progress and
prospects, to appear in the Proceedings of the Seventh Marcel Grossman
Meeting on General Relativity, preprint available as gr-qc/9410018.

\bibitem{Blencowe} M.\ Blencowe, The Hamiltonian constraint in quantum
gravity, {\sl Nucl.\ Phys.\ }{\bf B341} (1990), 213-251.

\bibitem{Borissov} R.\ Borissov, Regularization of the Hamiltonian constraint
and closure of the constraint algebra, Report TU-94-11, preprint
available as gr-qc/9411038.

\bibitem{BP} B.\ Br\"ugmann and J.\ Pullin, On the constraints of quantum
gravity in the loop representation, {\sl Nucl.\ Phys.\ }{\bf B390} (1993),
399-438.

\bibitem{G} R.\ Gambini, Loop space representation of quantum general
relativity and the group of loops, {\sl Phys.\ Lett.\ }{\bf B255} (1991),
180-188.

\bibitem{JacSmo} T.\ Jacobson and L.\ Smolin, Nonperturbative quantum
geometries, {\sl Nucl.\ Phys.\ }{\bf B299} (1988), 295-345.

\bibitem{Ash2} A.\ Ashtekar, Mathematical problems of
non-perturbative quantum general relativity, in {\sl Proceedings
of the 1992 Les Houches Summer School on Gravitation and
Quantization,} ed.\ B.\ Julia, North-Holland, Amsterdam, 1993.

\bibitem{TecRov} H.\ A.\ Morales-T\'ecotl and C.\ Rovelli, Fermions in quantum
gravity, {\sl Phys.\ Rev.\ Lett.\ }{\bf 72} (1994), 3642-3645.

\bibitem{carlip} S.\ Carlip, The statistical mechanics of the (2+1)-dimensional
black hole, Report UCD-94-32, preprint available as gr-qc/9409052.

\bibitem{BCM} A.\ Balachandran, L.\ Chandar, and A.\ Momen,
Edge states in gravity and black hole physics, Report SU-4240-590,
preprint available as gr-qc/9412019.

\bibitem{RS}  C.\ Rovelli and L.\ Smolin, Discreteness of area and volume  in
quantum gravity, Report CGPG-94/11/1, preprint available as gr-qc/9411005.

\bibitem{HH} S.\ Hawking and G.\ Horowitz, The gravitational Hamiltonian
action, entropy and surface terms, Report DAMTP/R 94-52 and UCSBTH-94-37,
preprint available as gr-qc/9501014.

\end{thebibliography}
